\documentclass[twocolumn]{revtex4-1}
\usepackage{graphicx}
\usepackage{latexsym}
\usepackage{amsmath}
\usepackage{bm}
\usepackage{amssymb}

\newcommand{\be}{\begin{equation}}
\newcommand{\ee}{\end{equation}}
\newcommand{\bea}{\begin{eqnarray}}
\newcommand{\eea}{\end{eqnarray}}
\begin{document}
\title{A linear response relation for perturbed Einstein's equations with a
Langevin source : Applications to perturbations in compact stars} 
\author{Seema Satin}
\affiliation{Indian Institute for Science Education and Research, Pune, India}
\email{seemasatin74@gmail.com}
\begin{abstract}
A new linear response relation for the perturbed Einstein's
 equation is introduced.  We give the idea of
 considering the metric perturbations as a linear response to fluid (matter)
 perturbations in strong gravity regions. This can be meaningful when the
perturbations in the system are driven by sources internal to the
fluid (matter) in the relativistic star.  The  aim is to study the strong
 regions with compact matter like that of  the internal  
structure of relativistic stars at sub-hydro mesoscopic scales with this new 
framework. 
The formulations are specifically done to address the generalized stochastic
perturbations which can arise in the  dense matter. These internally sourced 
perturbations lead to the possibility of equilibrium and non-equilibrium 
(dynamical or thermal)  statistical analysis for the properties 
of compact matter at the sub-hydro mesoscopic scales. 
A general relativistic Langevin formalism, defining a  
 random driving source and  its analytical solutions for a simple example 
are given. 
With a first principles approach this new  framework and its potential towards 
building up a theme of research in asteroseismology is discussed. 
\end{abstract}
\maketitle
\section{Introduction}
In this article we give a basic new theoretical formulation for
the perturbed Einstein's equation without violating the principles of general
 relativity. This is done by incorporating a linear
response relation between the metric and fluid  perturbations  in  linearized
first order perturbed Einstein's field equations. Such an idea  emanates
from the interest in exploring the nature of compact 
exotic matter that the interiors of relativistic stars contain.  Though there 
are several approaches in asteroseimology with active
research in this direction, they are all based on either the astroparticle
 and nuclear physics details or on the theoretical
 framework of the
 perturbed Einstein's equations given in the form of 
\be  \label{eq:perteq}
\delta G_{ab}(x)[h] = 8 \pi \delta T_{ab}(x)[h,\xi ]
\ee
where the background metric is $g_{ab}$ and its perturbations are given by 
$h_{ab}$, while $\xi_a$ denotes the displacement vector of the fluid 
trajectories. 
The perturbing agency whose details are considered insignificant and ignored 
often lies outside the configuration of interest.
The fact that sources of perturbations  may  lie inside the compact objects,
has  not  been considered or modelled yet as we do here in the theoretical 
framework of stochastic gravity. 
Hence observational consequences or classification for such specific 
perturbations has also not been yet explored.
If one is to consider any sources which are interior to the relativistic 
star and can perturb the fluid system, then the model described by the 
perturbed Einstein's equations should include them in principle. 
  
 The observational interests of 
the astronomers find connections with the available and well established 
theoretical framework that is understood in terms of various signals 
which can be received/filtered out from the  different kind of radiations,
 upto possible detectable precision.  
The theoretical base for asteroseimology was formed by developing the
framework of linearized first order perturbations of the 
Einstein's equations \cite{friedschutz,schutz1}.
The analysis of such perturbations in compact stars begins with the
study of stability and equilibrium properties \cite{friedschutz,fried1}
of the spherically symmetric star models where radial perturbations have their
significance. Radial perturbations also have a role to play in areas of 
complete gravitational collapse  \cite{rwald, joshi} for dust and fluid models
 of relativistic stars and the cosmic censorship conjecture investigations.
For the gravitational waves asteroseismology \cite{kokkotas,nils1, Schutz},
 the interest lies in non-radial perturbations.
All the above directions are developed in order to understand the interiors
 and dynamical properties of these compact objects and the nature of the exotic
 matter that they are composed of.
  Specifically the mode analysis of the
 perturbations characterize the internal structure of the massive stars.
This has gained more importance due to recent achievements in gravitational
wave detection \cite{abbot,abbot1}. 
The perturbations are considered as deterministic shifts in the trajectories 
of the fluid and the spacetime metric. For the equilibrium configurations they
have an oscillating form with the possibility of  growth or decay.
The evolution (growth or decay) of such oscillations is studied using the modes
analysis of the frequency spectrum. 
The characteristic feature of the  induced perturbations which  we are 
interested in are different and have a random generalized stochastic nature
as opposed to the usual deterministic perturbations considered 
 in asteroseismology. We use the concept of
generalized stochasticity for a spacetime structure, introduced in
\cite{seemacorr} for our formulations.
The significance of studying such random perturbations induced by sources
internal to the massive objects becomes clearer as we proceed 
further in all details, and sum up in the concluding
section of this article. 

The background metric
and fluid  remain exactly same and untouched by these new formulations and 
 retain the deterministic nature. Hence the background  unperturbed
metric $g_{ab}$ or fluid variables like pressure , energy density etc still 
remain definite deterministic quantities. 
  
We also model and characterize the generalized stochastic perturbations with
 an aim to probe 
through new scales of interest inside the dense matter of such objects.
While astroparticle and nuclear physics have seen elaborate developments and are
active areas in research \cite{schmidt,quark,pano}, there is scope to
 understand more about how quantum matter as bulk
behaves in strong gravity regions inside these compact objects. As part of
developing a new perspective, we  attempt to explore the  intermediate length
 scales inside the massive stars. This can be carried out by defining
and obtaining expressions for  extended (point separated correlations of) 
fluid properties which can be interesting and are likely to show deviation 
from the local 
properties of relativistic fluids. The relativistic fluid
modelling for compact matter stars is an an active area of research 
\cite{nils,kovtun}, and 
emphasis is laid on modelling the stress energy 
tensor, considering a flat spacetime background for the local properties at
a given point. We would call the new intermediate scales that we are
 interested in, as the sub-hydro mesoscopic scales, as they are 
expected to lie somewhat below the macroscopic classical 
hydro - static/dynamic length 
scales but above the microscopic scales (quantum or classical).  
For quantum or classical microscopic scales the effect of gravity is 
negligible on the localized particles that compose the dense matter. It is only 
for the extended properties of the 
 bulk  that one may be able to associate the effect of gravity on the 
dense exotic matter. An important  question  to ask here is, at which 
length scales in the system of the bulk matter do the effects of gravity 
start showing up minutely at least, and in which way can they be manifested.
Such questions arise in addition to the other  studies over the properties
 of dense matter that are being carried out presently.  
This is certainly a very interesting open question for which the answers would 
be specific to the massivenss of the astrophysical object and the inherent
characteristics of exotic matter at various depths in different types 
of compact objects. It is through such investigations that our 
understanding over nature of dense matter in strong gravity may be benefitted. 
With such an overall motivation we begin the new formalism in  this article.
 
As the first example of matter content, we would enhance the cold dense
matter with (non-thermal) random sources originating in the bulk.
 The 
microscopic details are not needed to capture a coarse grained description in
terms of fluid variables. Still a physical motivation for the
coarse grained effects is indeed of fundamental importance. 
Such sources may arise due to various  mechanical (quantum or classical ) 
or thermal reasons as in any physical system. In this article
 we give a particular source with a physical picture for spherically symmetric
relativistic star,  which
can  give rise to statistical  perturbations in the astrophysical body. But 
firstly, we begin with the new linear response relation for general 
relativistic applications.  
\section{A Linear Response Relation for perturbations in relativistic stars}
\label{sec:1} 
A linear response theory (LRT) for dynamical systems is based on the relation
between an external perturbing agency and its linear response on the 
coupled system \cite{volterra,cond,kubo}. Such a theory, which may not
 necessarily be based on considering a Hamiltonian for a classical system, 
 is valid  for small disturbances near
dynamical/thermal equilibrium states of the system. This is a general 
principle which acts as a base
for a vast number of theoretical developments in physics and natural sciences.
The applications of the LRT spans over thermodynamics, mechanical systems, 
biological systems, electrical circuits and chemical sciences. 
The linear response theory is based on the general principle where the
response (output) of a system ( say denoted by y(t) for a position) depends on
 some input $j(t) $, such that 
\be \label{eq:ltr1}
y(t) = \int_{-\infty}^t dt' \chi(t-t') j(t') + ...
\ee
Approximately, $y(t) $, is a weighted sum of the previous values of $ j(t')$, 
with the weigths given by the linear response $\chi(t-t') $. The term on the
rhs of this equation is the leading order term of a Volterra expansion
for the full non-linear theory \cite{volterra}. If the system is highly 
non-linear, higher order terms on the rhs become important.  
In its simplest form, linear response function describes the 
input output relationship, e.g a signal transducer such as a radio
turning electromagnetic waves into music, or a neuron turning synaptic  
into response.
The Kubo formalism for linear reponse theory \cite{kubo} is developed in the
 context of quantum statistics, which considers the force $j(t)$ as a
 perturbation of
the Hamiltonian, while the output $y(t) $ is the perturbation of the 
thermal expectation of another measurable quantity. The Hamiltonian 
formulation is generally used for quantum systems. For classical systems a
 Hamiltonian formalism is not necessary for a linear reponse relation and 
can be stated in a simpler way  in terms of any physical variable of the system
on which a response of a perturbing agency can be felt in a linear way. 

Specific developments of the linear response theory use the general
principle and formulate  details according to the required
 applications. This takes into account
the relevant variables of interest and a division between degrees of
freedom which suits appropriately in terms of system and environment. The
system-environment division is a general way to be able to frame consistently
the perturbing agency (environment) as the degrees of freedom  
that can be separated from the system which responds in a characteristic way
to the perturbation. Few  examples where a system-environment division 
is useful are, quantum open systems and theory of fluctuations in statistical
physics \cite{nobel,lifshitz} being amongst the most active areas currently.
 Based on quantum open systems, the theory of stochastic
semiclassical gravity \cite{bei}  which aims at studying structure formation in
the early universe cosmology is being developed. This theory
classifies metric perturbations as the system and quantum matter fields
as enviorment. Though the approach for semiclassical stochastic gravity
is based on this division, it follows a different course than the LRT (linear 
response theory), and uses the CTP (closed time path) and influence action
 formalism for its basic developments. 
It is indeed the case that the LRT is not necessarily limited
to stochastic fluctuations or statistical physics, though it forms a
cornerstone in statistical physics, but extends to
deterministic perturbations in dynamical systems \cite{volterra} and is a more
 general principle. 

For our developments  regarding the perturbations of relativistic stars
composed of dense matter, (and a study of classical perturbations as in 
asteroseismolgy)
we borrow the view of the perturbations of matter fields (dense fluid)
  as environment and the perturbations of spacetime geometry (metric) as the
 system. We propose the
  linear response relation in terms of  the metric perturbations
to disturbances in the fluid dense matter of the compact object using the
 principle of LRT for the relativistic star configurations.  
Such a division is easily possible with the fundamental elements of 
perturbations as  $h_{ab}$ for the spacetime metric  and $\xi_a$  for fluid
matter in a linear theory.  These fundamental variables have been widely used 
in all developments in asteroseismology as deterministic quantities.
Thus a linear response relation  suitable for the configuration of a perturbed 
relativistic star can be given by,  
\be
F[h;x) = \int K(x-x') f[\xi,x') d^4x'
\ee
for  a $ 3+1 $ dimensional spacetime, where $F[h;x)$ and $f[\xi,x)$ denote 
 perturbed quantities in terms of the metric perturbations denoted by $h$ and 
the fluid displacement vector denoted
by $\xi$. The kernel $ K(x-x') $ connects the two and decides the reponse.  
The kernel as proposed here is for a spacetime structure hence it carries
the weight of the previous times as well as a spatial range which
contributes to the weighted sum of the input. This general form of 
the linear response relation is thus an extension of equation (\ref{eq:ltr1}) 
suitable for the strong
 gravity regions governed by the Einstein's equations and having a spacetime
 structure. Similar to equation (\ref{eq:ltr1}), it will have higher order 
terms for nonlinear perturbations  and give a further scope for
interesting developments in the strong gravity regions. However in this 
article as a first step, and with the aim to analyse the dense matter star 
perturbations for asteroseismological considerations, we restrict ourselves to
the linear theory. This step is also needed to establish the first 
principles correctly. 
  
Writing the above linear response relation in terms of the perturbed 
Einstein's equations gives it the explicit form 
\be \label{eq:eltr}
\delta G_{ab}[h;x) - 8 \pi  \delta T_{ab}[h;x) = 8 \pi \int K(x-x') \delta
T_{ab}[\xi,x') d^4 x'
\ee
The perturbations on the l.h.s are a linear response
to the fluid trajectory being displaced, the cause/source of which may lie
outside or inside the gravitating body. We model this  
 with the condition, that a  $\delta$- correlation of the response
kernel $K(x-x') = \delta^4(x-x') $ leads to the usual perturbed Einstein's
equations (\ref{eq:perteq}).
In this case, we can assume that the source of perturbations lies outside the
relativistic star. 
If an internal noise/source (that we show below) plays a role in 
perturbing the trajectories of fluid, the  model of such a noise may also have 
a few zero or vanishing components for the Einstein's equations due to 
physical reasons. 
Then for those components in the above equation it is appropriate to
take a  delta correlated response function. The latter
aspect is important for our considerations as the background noise may
have zero and non-zero components for a tensor that describes it. This 
would become clearer as we proceed on the specific model and do the first
exercise in the following sections.  These mathematical formulations guide
us through the new theoretical way of analysis. Similarly the theoretical
ideas which we borrow and extend from other areas in physics decide the
mathematical formulations for the astrophysical systems on a spacetime 
geometry. 
For the sources lying outside the astronomical object, the case reduces to
the regular developments in asterosiemology \cite{fried1,kokkotas}.  Then
our linear response relation above reduces very easily to the regular
perturbed Einstein's equations, and hence shows consistency. 
\subsection{The complete form for the classical Einstein Langevin equation} 
Inspired by the semiclassical Einstein Langevin equation  \cite{bei}, 
for an analogous development for the classical case,  one can add a source
 term $\tau_{ab}(x)[g;x) $ to the perturbed Einstein's equations
 (\ref{eq:eltr}). Such a prescription phenomenologically describes 
perturbations to the Einstein's equations, which arise due to internal
sources in the system.  We model these as the generalized fluctuations of 
matter fields, which are classical in nature as
opposed to the quantum fluctuations of matter fields in the theory 
of semiclassical
stochastic gravity. Hence they carry a different nature altogether, and 
are defined in terms of a Langevin noise in the classical matter fields.
The LTR and the perturbed equations do not describe these sources, but they 
describe the perturbations that are induced by the Langevin term.  

Thus to complete the new formalism which is the aim here , we add a source 
term (having origin inside the bulk of dense matter) defined on the background 
spacetime $g_{ab}$.
 For the above equation this consistently takes a full form as an internally 
sourced perturbed Einstein's equation, by adding $\tau_{ab}[g;x) $ to
(\ref{eq:eltr}), as    
\begin{widetext}
\be \label{eq:elfull}
\delta G_{ab}[h;x) - 8 \pi  \delta T_{ab}[h;x)-  8 \pi \int K(x-x') \delta
T_{ab}[\xi,x') d^4 x' =  \tau_{ab}[g,x)
\ee
\end{widetext}
The Einstein's equation  $G_{ab}[g;x) - 8 \pi T_{ab}[g;x)=0 $ consistently
 balances the unperturbed configuration, while the generalized stochastic
 source $\tau_{ab}[g;x)$ in the background is balanced  by the perturbations
it induced in the configuration.
 Note that $\tau_{ab} $ is defined on the background spacetime
$g_{ab}$, and not on the perturbed metric. 
The above equation is linear in perturbations and hence covariantly conserved
with respect to the background metric $g_{ab}$.  One usually assumes this 
for the linear first order perturbations in general relativity. Thus 
$ \nabla_a \tau^{ab} [g,x) = 0 $ is essentially to be satisfied for the source
term, and hence it is a covariantly conserved term in the equation. 
It is necessary at this point to bring attention to the fact, and for clarity 
of the work done in this article, that the induced perturbations and
the stochastic noise are two different entities here, though they are 
obtained from the same form of the stress energy tensor. This is in 
analogy with the semiclassical Einstein-Langevin equation \cite{bei}, where
 the quantum noise is obtained from the qunatum stress-energy tensor.   
The noise or source
in the system may be added in any consistent way, but with the matter 
fields under consideration, a
 describtion  of the sources in terms of generalized fluctuations of 
the stress energy tensor is appropriate. This is then  also consistent with the 
Einstein's equations and the basic principles of general relativity. As in the
 case of 
semiclassical stochastic gravity, for the classical case also, such a noise
 term has to be covariantly
conserved, and obtained in an elaborate form. This is unlike the 
gaussian white noise for the Brownian motion or other simpler cases in 
Newtonian systems which can be written in a handwaving way. The elaborate
form for the source term hence has to be obtained from the specific
matter fields with the underlying spacetime metric on which it is defined
and is non-trivial, though we add  it to the Einstein's equations in a
phenomenological way. 
Details of the generalized random source as introduced in \cite{seemacorr},
 which extend the concept of randomness in physical variables w.r.t the 
spacetime coordinates, rather than just the temporal coordinate, can be 
modelled this way.  
We thus define the generalized  Langevin source as
\be
 < \tau_{ab}(x)> = 0 , <\tau_{ab}(x) \tau_{c'd'}(x') > = N_{abc'd'}(x,x')
\ee
where $ < ... > $ denotes the statistical average and
 the two point noise kernel $N_{abc'd'}(x,x')$ is important as we will
see later. 

We consider the generalized stochastic source 
in the form of $\tau_{ab}(x) = \delta_s T_{ab}(x)$.
The 's' in $\delta_s$  labels it as the fluid "source" in the unperturbed 
background $g_{ab}$. This is different from the perturbations denoted by 
$ \delta T_{ab} (x) $ due to shift in trajectories of the fluid given by
 $\xi$ and $h_{ab}$ for the metric.  We give an explicit form of the 
generalized fluctuations of fluid variables as components of $\tau_{ab}$ 
for the example worked out in this article, in a later section.   
\section{The spherically symmetric relativistic star model with
generalized random effects in dense matter fluids} 
The astrophysical configuration that we consider in this article
 to explain and show our framework is detailed in the subsections
below. The work done in this article involves  a  rigourous formulation for the
classical E-L (Einstein-Langevin) equation using the linear response relation,
 which was not done in the previous references 
that proposed the basic idea \cite{seema1,seema2}. Though the final solutions
for perturbations in this article look similar to those given in the above
references, this article is more than incremental over the previous 
ones at the conceptual and theoretical level. One can also see that the final
 results here contain the response kernel along with the slightly different
 form of expressions for the perturbations  with  a random  oscillatory part
 as well.   
\subsection{Model of spherically symmetric relativistic star}  \label{subsec:1}
We consider a model of spherically symmetric star with an initial phase of a 
dynamical collapse. The metric for this in a  comoving  frame is given 
by
\be \label{eq:com}
ds^2 = - e^{2 \nu(r,t)} dt^2 + e^{2 \lambda(r,t) } dr^2 + R^2(r,t) d \Omega^2
\ee
We are interested in an analysis when such a radial collapse reaches a 
dynamical near-equilibrium state. This happens when the degeneracy pressure 
of the matter can counteract the gravitational pressure and sustain the
structure at some point in the evolution. These stages are
decided by the mass of the star. 
 We do not touch upon  the black hole singularity  end state of collapse
with our formalism, as this is
 restricted to the objects with dense matter fields. In
the current literature for relativistic star interiors, several states of 
quantum matter are considered with a nuclear physics and astroparticle physics
 description \cite{schmidt,quark}. We do not progress using these directions,
 but add from the very basics a different approach without
starting from the microscopic nuclear physics assumptions and theory. 
 
We proceed with the  overall description in terms of the bulk 
properties of dense fluids with an  intermediate scale picture 
 using a  perturbative approach. From quantum physics the ingredients in terms 
of  the "degeneracy pressure" in cold dense matter which plays a major role at
 global scales in balancing/opposing the gravitational pressure 
are carried along and suffice for the purpose here.  
 We consider the evolution of the star, towards the end stages of
dynamical collapse, such that one can study the configuration around
the static equilibrium in a linear perturbative way.  Such studies are
usual for oscillating stars and their radial perturbations, the perturbations
in the oscillating star are dynamical in nature while the equilibrium
background can be considered static. Here we would like to emphasise that
we are not considering a static relativisitic star and introducing any 
perturbations in the static fluid, but the time evolution rather goes
from dynamical to a later static equilibrium stage. However, it is around the 
equilibrium as for any perturbative studies of a physical system, that
we expand our equations. Thus the equilibirum metric $g_{ab}(x) $ is a static 
spherically symmetric spacetime and also is deterministic in nature, while its
perturbations $h_{ab}(x) $ have a time dependence and are dynamical. 
We thus have the perturbed configuration as $g_{ab}(x) + < h_{ab}(x)> $
where $h_{ab}(x) $ which is generalized stochastic in nature as we will see 
and hence in our case has meaning only as a stochastic average will be
vanishing $ <h_{ab}(x)> =0 $. However the two point correlations are 
non-vanishing
and build up the statistical theory for these stochastic perturbations.
 The important point to stess here is that, the background around which these
linear dynamical perturbtaions take place is taken to be equilibrium state  
which is given  Schwarzchild coordinates as, 
\be \label{eq:spher}
ds^2 = - e^{2 \nu(r)} dt^2 + e^{2 \lambda(r) } dr^2 + r^2 d \Omega^2
\ee
The stress-energy tensor for a perfect fluid is given by

\be \label{eq:stress1}
T_{ab} = (\epsilon+ p) u_a u_b + g_{ab} p
\ee
where the four-velocity $u^a = e^{-\nu}(1,0,0,0)$.
\subsection{Model of matter fields with general stochastic effects}
The model of matter fields in a relativistic star as given by the stress 
energy tensor in the subsection above can be enhanced with random fluctuations
as we show below even for the perfect fluid approximation.  No thermal
effects are considered here, as we restrict to the perfect fluid approximation
for the work presented in this article. 
\subsubsection{Preliminaries of the effective fluid approximation
in compact dense matter stars under the effect of gravity} 
The pressure in ordinary fluids  is due to the equilibrium  thermal properties
and related to the flow of fluid defined by the velocity vector. This  has a
 different origin than in the effective fluid models for dense matter.
The effective fluid model of the dense matter stars  is different than
 the regular fluids in that, the pressure in the former case arises due to the
degeneracy of the quantum matter in the bulk. 
 Though for the hot dense matter, thermal
pressure  and temperature play a similar role as for ordinary fluids with
an equilvalent thermodyamics of the hot dense matter, there are more
involved things that make up their inherent properties. The stress
 energy tensor  defines all the effective fluid properites for
the hot as well as cold dense matter in  a  spacetime structure. Specifically
for the cold dense matter stars, the degeneracy pressure  
is responsible for counteracting the strong gravity which holds the shells
together and prevents further collapse. The energy density is related by an
 equation
 of state with the pressure and has a similar origin. The thermal association
 of pressure in ordinary fluids is thus replaced by the interplay of gravity
 and quantum degeneracy of matter  in the massive astrophysical
objects.  We use this feature in our work to explore non-thermal
equilibrium and non-equilibrium statistical properties of the dense matter.
An important aspect is that, here, the four-velocity of the effective fluid 
being a kinematic term does not
 arise due to the degeneracy pressure of quantum fluid. This is unlike the 
ordinary fluids where thermal pressure and thermal velocity have a 
direct connections. 
\subsubsection{Origin of a driving source in cold dense matter stars}
It is known that at  certain stages in the evolution, the degeneracy 
pressure of the particles that constitute the matter at various
depths, starts opposing the gravitational pressure and hence counteracting
the radial collapse. The picture of a smooth collapse where an exact static
equilibrium is attained if the gravitational pressure balances the
degeneracy pressure of the species of particles is an idealized one. 
We consider a more refined picture of  near stable as 
well as unstable  equilibrium (dynamical) states. The unstable
equilibrium state can occur intermittently, when the evolution  progresses
 over a span of stages starting from electron degeneracy, neutron degeneracy 
and further on for more elementary particle states.
Thus depending on the  total mass of the gravitating cloud, the collapse
could stop at one stage  or have a short unstable equilibrium/ near
equilibrium phase before evolving further. At such an equilibrium  
 the degeneracy pressure of the corresponding particles and the gravitational
pressure just balance each other. For the spherically symmetric star one can
  view this in terms of  the radial shells of the star.
For such (dynamical) near-equilibrium states  we would be interested in 
modelling the trigger sources in matter which can act as seeds to perturb the 
system.   

Consider a spherically symmetric relativistic star which has burned its fuel and
has collapsed to a near equilibrium cold dense matter, where thermal effects
are negligible.
Then equation (\ref{eq:com} ) describes in comoving coordinates the radial
collapse, which in the Schwarzchild coordinates takes the form 
\be
ds^2 = - e^{2 \nu(t,r)} dt^2 + e^{2 \lambda(t,r)} dr^2 + r^2  d\Omega
\ee
This represents a spherically symmetric dynamical collapse of the relativistic
star. The final stable state is reached when the collapse stops and the 
configuarion takes the form given by (\ref{eq:spher}). We address the near 
equilibrium state of the star such that, one can linearize the equations
around the equilibrium configuraion given by (\ref{eq:spher}). Thus the
background that we assume is a static case,  around which we have first order
linear perturbations. The perturbations around the 
equilibrium carry a dynamical nature.    

The phenomena of balance at equilibrium points during the evolution
in cold
dense matter stars, will certainly be due to the global nature ( mass ) of
the dense matter and the inherent bulk degeneracy pressure of  shells
compressing each other as layers in a spherically symmetic configuration. Thus 
one can view the phenomena of balancing $ p_g(r,t) = p_d(r,t), $ after a
certain at a certain $t=t_0$ when the configurations reaches a near
 equilibrium state.
This mechanism can set small randomized oscillations at sub-hydro mesoscopic 
scales in the value of pressure of the fluid
for a radial shell at $r$, in order to balance and sustain the system 
consistently, thus  the condition  $\delta_s p_g(r,t) = \delta_s p_d(r,t)$
 follows. 
For the near equilibrium stage these random (in spacetime) radial oscillations
 are expected to maintaining the spherical symmetry.
 One can also model deviations from the spherical symmetry, but we would address
 such cases in later articles to follow. Here we take the simplest  form and
consistently put  $ \delta_s p_g(r) e^{i \omega_rt} = \delta_s p_d(r)
 e^{i \omega_r t}$. The  small amplitude pressure seeds  given by 
$\delta_s p(r)$  originate due to the mechanism of balancing the two
 pressures $p_d(r,t) $ and $p_g(r,t) $. 
 Thus $\delta_s p(r) $ has a probability distribution 
$ P(\delta_s p(r)) $ and only a statistical average $ < \delta_s p(r) >$ 
can be meaningfully used for any physical interpretation. Similarly, the  
oscillatory part can be modelled to have  a random frequency $\omega_r$
 with a distribution $P(\omega_r)$.
This can happen at various depths around the different  (unstable) 
equilibrium stages that the relativstic star may go through in the 
course of its evolution. 
The frequency $\omega_r$ of these random oscillations then can 
capture/characterize features which can be enhanced by the perturbations that 
are
induced in the fluid, as we show later.  The induced random perturbations can
 have a  growth or decay in a dense fluid which  is independent of its source.
 Hence we include the whole range of theoretically possible random oscillations
as described above, as the mechanical sources for the model considered here.  
\subsubsection{ A  model of generalized stochastic  source in spherically 
symmetric relativistic star of cold dense matter }   
For a spherically symmetric configuration, it turns out that the non-zero 
components of $\tau_{ab}(x)$ in the cold ideal fluid by definition take the
form, 
\bea  \label{eq:source}
& & \tau^t_t(r) e^{i \omega_r  t} = \delta_s \epsilon (r) e^{ i \omega_r t}
\nonumber \\ 
& & \tau^r_r(r) e^{i \omega_r  t} = \tau^\theta_\theta (r) e^{i \omega_r t} = 
\tau^\phi_\phi (r) e^{i \omega_r t } = \delta_s p (r) e^{ i \omega_r t}
\eea
With  $\delta_s p(r) $ and $\delta_s \epsilon(r)$ denoting random amplitudes
 at different radii and  the  oscillatory parts with random
frequency  $\omega_r$ having a distribution density.  
 
 The the pressure and energy
 density in dense matter are independent of the 3- velocity of any observer
and characterize the nature of cold dense degenerate matter. 
Since we address the cold matter ideal fluid in this article, one can rule out
 any thermal contributions to the pressure. 
Therefore we do not consider any contributions of the  3- velocity
fluctuations in the sources around the
equilibirum, though as we will see later, that the perturbation to the 
3-velocity 
in terms of the Lagrangian displacement vector for fluid trajectories is 
 taken into account consistently, in our work. 
Hence only the contributions from bulk properties like pressure and energy 
density at different radial layers
 inside the star act as Langevin sources for this configuration. 
 We will rather show how a non-zero value of perturbed radial 3-velocity can be
 induced by these sources.
The necessary property that this model of noise of Langevin
term has to satisfy is the covariant conservation w.r.t the background
metric $\nabla_a \tau^a_b = 0 $. Given that we are taking 
the unperturbed $g_{ab} $ as the background around which the system is 
perturbed, this necessarily calls for covariant conservation of only
the amplitudes of  $\tau(x)$ and not the oscillatory part. This has also
been discussed in \cite{seema1}. Hence the condition $\nabla_a \tau^{ab}(r)
=0 $ is satified by the model of noise.  This  is seen to arise
due to the fact that we are perturbing around a static equilibrium metric
 and  also that the first order perturbations to Einstein's tensor $\delta
G_{ab}(x) $ and the stress tensor $ \delta T_{ab}(x) $  are 
covariantly conserved w.r.t the background spacetime is an approximation 
for small perturbation.
Thus $e^{i \omega_r t} $ is an ad-hoc for the covariance conservation
property  w.r.t the background, but there is no discrepancy with respect to
the balancing of the r.h.s and l.h.s  in the Einstein-Langevin equation
(\ref{eq:elfull}) due to this.
Thus for the above non-zero components of the generalized stochastic source, 
the conditons that the amplitudes of the stochastic $\tau_{ab}(r) $
sources should satisfy, given by the condition $\nabla_a \tau^{ab}(r) = 0 $ ,
 gives
\be
\delta_s p(r)' + \nu' (\delta_s \epsilon(r) + \delta_s p(r)) =0
\ee 
which shows the valid condition for the  amplitudes $\delta_s p(r)$ and
$\delta_s \epsilon(r)$ of the sources ( which we have assumed also have a 
random nature w.r.t the the radial coordniate giving roughness in pressure and
energy density as a new physical insight). As we have probed with 
such a formulation, scales that are smaller than the hydrodynamic macro 
scales, we can expect to see such effects in the dense fluid, which have not 
been modelled yet. This is also different than the nuclear physics
and quantum details as we have already mentioned, and carry us into a new
 sub-hydro regime in which 
one can talk of such effects around the equilibrium fluid model, in  terms of 
the  fluid variables.   
To acertain that the term $e^{i \omega_r t} $ introduced phenomenologically
for the noise is not wrong, one can compare the picture at these mesoscopic
 scales by
taking time averages of these seeds of fluctuations which just give
the averaged out radial amplitude. One can view this similar to the thermal
noise in a system, where  the randomness in the
four velocity of particles at microscopic scales, can be averaged out and 
one talks in terms of average four -velocity at larger length scales of the
fluid particles. Similarly here one can
consider the time averaged stochastic source terms such that for reasons of
validity for the background spacetime and general covariance of the noise,
the factor $e^{i \omega_r t}$ in the term $\tau_{ab}$ does not raise an issue, 
because we are going at more refined scales. 
\section{The induced perturbations in the system}  \label{sec:4}
 Beginning with the Einstein-Langevin equation (\ref{eq:elfull}), as
\begin{widetext}
\be \label{eq:elf} 
\delta G_{ab}[h;x) - 8 \pi  \delta T_{ab}[h;x)-  8 \pi \int K(x-x') \delta
T_{ab}[\xi,x') d^4 x' =  \tau_{ab}[g,x) 
\ee
\end{widetext}
we show how the background source $\tau_{ab}[g;x)$ induces perturbations in the
system
For near equilibrium states, we expect the spherical symmetry to be 
maintained. Hence for the perturbations that preserve the spherical symmetry of 
the star, namely the radial perturbations the  above equation  can be solved
using the perturbed components. 
subsection{Perturbations around the static equilibrium}
The perturbed system around the static equilibrium is given in term of the
pertubred $\xi $ Lagrangian displacement and the metric potentials
$\delta \lambda$ and $\delta \nu $.  How are these perturbations introduced
is a different question, which we address later in our system. However such
perturbations in relativistic stars are well studies  in literature and
are known to be induced by some perturbating agency which is no
specified in the equations, as it is considered unimportant to characterise
the perturbations of the system. Often such an agency is assumed to lie outside the gravitating body and  hence ignored.  The perturbed equations maintain
the usual form, except for our case we have given an Einstein Langevin  form
with a linear response relations.

The  perturbations of the fluid stress
 tensor $T_{ab}(x)$ has two parts, one corresponding to the fluid
displacement vector $\xi_a$ and other to the metric perturbations $h_{ab}$.
\bea
\delta u^a & = & q^a_b \mathcal{L}_u \xi^b + \frac{1}{2} u^a u^c u^d h_{cd}
\mbox{ (where $q^a_b = u^a u_b + \delta^a_b $) }  \nonumber \\
\delta \epsilon & = &  - \frac{1}{2} ( \epsilon + p) q^{ab}( h_{ab} + \nabla_a
\xi_b + \nabla_b \xi_a ) - \bm{\xi}\cdot \nabla \epsilon \nonumber \\
\delta p & = & -  \frac{1}{2} \Gamma_1 p q^{ab}( h_{ab} + \nabla_a
\xi_b + \nabla_b \xi_a ) - \bm{\xi}\cdot \nabla p
\eea
where $\Gamma_1 = \frac{\epsilon+ p}{p} \frac{dp}{d\epsilon}$ is the adiabatic
index for the relativistic star.
For the radial perturbations that we are interested in, $\xi_a$ has only one
 non-zero component $\xi_r(x)$. The above
 separated into $\xi_r \equiv \xi$ and metric
potentials $ \delta \lambda(r)$, $ \delta \nu(r) $  are given by
\bea
\delta p[\xi] & = & - \Gamma_1 p\frac{e^{-\lambda}}{r^2} [ e^{\lambda} r^2
\xi ]' - \xi p' \\
\delta \epsilon [\xi] & = & - (p + \epsilon)
\frac{ e^{-\lambda}}{r^2} [ e^{\lambda} r^2 \xi]' - \xi \epsilon' \\
\delta u^r[\xi] & = &  \mathcal{L}_u \xi^r = e^{-\nu} \dot{\xi}
\eea
\bea
\delta p[h] & = & - \Gamma_1 p \frac{e^{-\lambda}}{r^2} \delta 
\lambda \\ 
\delta \epsilon [h] & = & - (p + \epsilon) \frac{e^{-\lambda(r)}}{r^2} 
\delta \lambda(r)  
\eea
Only the $t-t$, $t-r$, $r-r$, $\theta-\theta$ , $\phi- \phi$
components of equation (\ref{eq:elf}) are non-zero for the case.
We work out such induced perturbations in the system by solving the
Einstein-Langevin equation with all the above information. Choosing for
simplicity $K(x-x') = \delta(r-r')K(t-t') $ which has only the radial and the
time dependence due to the symmetry of the system, the  
non-zero components of E-L equation are as following. 
The source term $\tau_{ab}[g_{ab},x) $ which is described in the previous
 section, is then responsible to feed these perturbations.   

The $t-t$ component for the Einstein-Langevin equation reads 
\begin{widetext}
\bea
& & \delta G^t_t [h](r,t) - 8 \pi \delta T^t_t [h] (r,t) - 8 \pi \int \delta 
T^t_t[\xi](r,t') K(t-t') dt' = \tau^t_t [g](r,t) \\
& & -2 e^{- 2 \lambda(r)} \frac{\delta \lambda(r,t)}{r} 
(\frac{1}{r} - \nu'(r) -\lambda(r))  - \frac{2}{r} \delta \lambda'(r,t)
  e^{-2 \lambda(r)} - 8 \pi \int K(t-t') [ (\epsilon(r)+ p(r) ) \{\xi'(r,t') +
(\lambda'(r) + \frac{2}{r}) \xi(r,t') \} \nonumber \\
& & + \epsilon'(r) \xi(r,t') ] dt'  =  8 \pi \delta_s \epsilon(r) 
e^{i {\omega}_r t}
\eea
\end{widetext}
 The induced fluid displacement $\xi$ can be assumed to be of a reasonable
 general form $\xi(r) e^{\gamma_r t} $, such that $\gamma_r $ is a complex
 number in general and the factor $e^{\gamma_r t} $ decides the evolution of
 the induced perturbations.    
Then the above equation takes the form,  
\begin{widetext}
\bea \label{eq:tt}
& & -2 e^{- 2 \lambda(r)} \frac{\delta \lambda(r,t)}{r}  
(\frac{1}{r} - \nu'(r) -\lambda(r))  - \frac{2}{r} \delta \lambda'(r,t) 
e^{-2 \lambda(r)} + 8 \pi \tilde{K}(\gamma_r) 
[ (\epsilon(r)+ p(r) ) \{\xi'(r,t) + (\lambda'(r) + \frac{2}{r}) \xi(r,t) \}
\nonumber \\
& & + \epsilon'(r) \xi(r,t) ] =  8 \pi \delta_s \epsilon(r) 
e^{i {\omega}_r t}
\eea
\end{widetext}
where $\tilde{K}(\gamma_r) $ is the Laplace transform of the response kernel
$K(t-t')$, taking $(t-t') = \bm{\tau} $ as we may consider a stationary local 
reponse (given by $r=r'$ and $\tau= t-t'$) for the induced perturbations, such
 that  
\bea
\int K(t-t') e^{- \gamma_r t'} dt' & =& -  e^{-  \gamma_r t} \int 
K(\mathbf{\tau})  e^{ - \gamma_r \bm{\tau}}  d \bm{\tau} \nonumber \\
& = & - e^{\gamma_r t } \tilde{K}(\gamma_r) 
\eea
The factor $\tilde{K}(\gamma_r)$ gives the susceptibility of the spacetime 
metric to get perturbed due to the fluid displacement building up,
and is specified by the $\gamma_r$ radial spectrum.
Similarly the $r-r$ component for equation (\ref{eq:elf}) gives, 
\begin{widetext}
\bea
& &  \delta G^r_r [h](r,t) - 8 \pi \delta T^r_r [h] (r,t) - 8 \pi \int \delta 
T^r_r[\xi](r,t') K(t-t')  dt' = \tau^r_r [g](r,t) \\
& & 2 e^{-2 \lambda(r)} [ \frac{ \delta \nu'(r,t)}{r} + ( \frac{1}{r^2} +
\frac{2 \nu'}{r} ) \delta \lambda(r,t)] + 8 \pi \Gamma_1 p \delta \lambda(r,t)
-  \nonumber \\
& & 8 \pi \Gamma_1 p  \tilde{K}(\gamma_r) [\xi'(r,t) + (\frac{2}{r} +
\lambda'(r) + \frac{\epsilon'(r)}{\epsilon(r) + p(r)}) \xi(r,t) ]  =
\delta_s p(r) e^{i {\omega}_r t } \label{eq:rr} 
\eea
For the $t-r $ component  of E-L equation 
\be
\delta G^t_r [h](r,t) - 8 \pi \delta T^t_r [h] (r,t) - 8 \pi \int \delta 
T^t_r[\xi](r,t') K(t-t') dt' = \tau^t_r [g](r,t)
\ee
\end{widetext}
With the model of the noise which we use here, $\tau^t_r(r,t) = 0 $ , as
there is no source term corresponding to $t-r$ component. As discussed earlier
 a delta correlation for the  response kernel in the components where the 
source term is vanishing can be used here. In the above equation
we can assume $K(t-t') = \delta (t-t')$. 
Also $\delta T^t_r [h](r,t) = 0 $ as $\delta u_r[h] =0 $ which is discussed
earlier, giving 
\be
- \frac{2}{r} e^{-2 \nu(r)} \dot{( \delta \lambda (r,t))} = 8 \pi   
e^{2 \lambda(r) - 2 \nu(r)} (\epsilon(r) + p(r) ) 
\dot{\xi}(r,t)  
\ee
where $\cdot$ denotes the time derivative. This gives the relation  
\be  \label{eq:xidel}
\xi(r,t) = - \frac{ \delta \lambda(r,t)}{(\nu'(r) + \lambda'(r))}
\ee
It  also follows that the $e^{\gamma_r t} $ dependence is same for
$\xi(r,t) $ and $\lambda(r,t)$. 
Though the $t-r$ component of the E-L equation seems to be source free,
it is not the case that the perturbations vanish here or occur due to any
source external to the system. For a complete picture the other components
of the E-L equation show how the induced perturbations are related to the
source term correctly. The $t-r$ component above is one of the set of
equations, and  shows a relation between $\delta \lambda $ and $\xi$, at
a given $(r,t)$. This is useful to obtain the explicit relation between the 
source terms and the perturbations in a simple and  direct way. 
Using equation (\ref{eq:xidel}) to solve equation (\ref{eq:tt}) the following
 can be easily obtained 
\bea
\delta \lambda (r,t) & =  & 4 \pi e^{\nu - \lambda} r \int_0^r 
\frac{e^{3 \lambda(r') - \nu(r')}}{[1 + \tilde{K}(\gamma_{r'})]} \delta_s 
\epsilon(r') e^{i {\omega}_{r'} t } dr'
\eea
Here, a concern about causality can be raised as we see the 
integration over the radial vector $r'$ spans over $[0,r]$, which means
that all the layers from the center of the star upto $r$ contribute 
to the perturbations  $\delta \lambda(r,t) $. This would need the information
from the contribution $\delta \epsilon(r') e^{i \omega_{r'} t} $ from
all radial shells starting from center to radius $r$ to travel
 instantaneously, which violates causality. However this can be 
taken care of by observing detials of  the factor $K(\gamma_r')$ on the rhs of
the above equation.
As it follows from the definition  that
$\tilde{K}(\gamma_r)  = \int K(\bm{\tau}) e^{- \gamma_r \bm{\tau}} d 
\bm{\tau} $, which 
is  essentially  limited to intervals $ \bm{\tau} = (t-t') $ only. Hence we
add a condition for the above  result/expression to be valid for 
$ r-r' \leq c(t-t') $ , $c$ being the speed of light signal. Hence the
valid integration limits can be modified  accordingly,   
\bea \label{eq:la}
\delta \lambda (r,t) & =  & 4 \pi e^{\nu - \lambda} r \int_{r_1}^r 
\frac{e^{3 \lambda(r') - \nu(r')}}{[1 + \tilde{K}(\gamma_{r'})]} \delta_s 
\epsilon(r') e^{i {\omega}_{r'} t } dr'
\eea
such that $r-r_1 \leq c \bm{\tau}$ where  $\tilde{K}(\gamma_{r'})$ is the
Laplace transfrom of $K(\bm{\tau})$  with interval $c \bm{\tau} $.     
The above  using the relation  (\ref{eq:xidel}) gives
\begin{widetext}
\bea \label{eq:xif}
\xi (r,t) & =  & 4 \pi (\nu'(r) + \lambda'(r)) e^{\nu - \lambda} r 
\int_{r_1}^r \frac{e^{3 \lambda(r') - \nu(r')}}{[1 + \tilde{K}(\gamma_{r'})]}
\delta_s \epsilon(r') e^{i {\omega}_{r'} t } dr'
\eea
Similarly from equation (\ref{eq:rr}) one gets, 
\bea \label{eq:nuf}
\delta \nu(r,t) & = & \int Y_1(r',\gamma_{r'}) ( \int \frac{ e^{ 3 
\lambda(r'') - \nu(r'')}}{[1 + \tilde{K}(\gamma_{r''}) ]} \delta_s 
\epsilon (r'') e^{i {\omega}_{r''} t } dr'') dr' + \int Y_2(r', \gamma_{r'})
\delta_s \epsilon ( r') e^{i {\omega}_{r'} t} dr' \nonumber \\
& & + \int 4 \pi r' e^{2 \lambda(r')} \delta_s p(r') e^{i {\omega}_{r'} t} dr'
\eea
\end{widetext}

The limits of intergration over $ r' $ and $r''$ would be such that the
causality conditions are not violated.
The factors,
\begin{widetext}
\bea
Y_1(r,\gamma_r) & = & - 4 \pi [ \tilde{K}(\gamma_r) C_s^2 \{ 2 e^{\nu - 
\lambda} + (1 + e^{\nu - \lambda}) r ( \nu' - \lambda') \} + ( \lambda'
+ \nu') C_s^2 r + 2 \nu' r + e^{\nu - \lambda} ] \\
Y_2(r,\gamma_r) & = & - 4 \pi r \frac{e^{2 \lambda(r)} \tilde{K}(\gamma_r)}{(1
+ \tilde{K}(\gamma_r))} C_s^2 
\eea
\end{widetext}
with $C_s = \sqrt{\frac{dp}{d\epsilon}} $, show the dependence on the speed 
of sound in dense matter. 
The appearance of the speed of sound in the above solutions, points to
the possibility of associating the issue of the disturbances in 
pressure and energy density propagating from shells separated by larger
 distances
inside the star and the causal effect to be related.
The speed of sound may decide upto which depth the contributions from the
sources can be taken. This may give a more stringent  criteria on the valid 
limits of integration $ [r_1,r] $ for the r.h.s  of equations  (\ref{eq:la}) 
(\ref{eq:xif}) and (\ref{eq:nuf}). 
 
The equations (\ref{eq:la}) (\ref{eq:xif}) (\ref{eq:nuf}) are the main 
expressions for obtaining the formal solutions of the Einstein-Langevin 
equation (\ref{eq:elf}) for the configuration considered in this article.
Characterizing such induced perturbations
and their mode analysis would be of interest to study their growth or decay in
the star configuration. A generalized stochastic nature 
even for the modes of random oscillations is the new feature 
corresponding to this type of perturbations that are induced due to internal
sources. However the progress on these lines would be the next stage 
of development for the theory, which would be carried out in furture work. 
The radial as well as non-radial perturbations are amenable to such a study,
aiming towards realistic configuartions of compact objects eventually.
One can see from equation (\ref{eq:la}) that the metric potential 
$\delta \lambda(r,t) $ is obtained by the integrated effect of the 
random source term $\delta_s \epsilon(r') e^{i \omega_r' t} $ 
over shells of radius $r'$  for an interval $[r_1 , r]$. Thus 
$\delta \lambda $ is  induced at the shell $r$ due to the 
integrated effect of the small amplitude energy density 
oscillations  over a certain radial depth in the star. We
see the appearance of suscpetibility of the spacetime $[1+K(\gamma_r)] $ 
on the rhs, which determines  how the spacetime metric or potentials
respond to these fluid energy density disturbances. The susceptibility 
$[1+K(\gamma_r)]$ is the "charactersistic appearing at the linear perturbative 
approximation scale"  of the spacetime geometry  on which 
the dense fluid is embedded . The factor $ K( \gamma_r) $  shows radial
 dependence for this case which says that the response felt by the metric 
perturbations has radial dependence in the spherically symmetric star.
We see similar factor of $K(\gamma_r)$ appearing in the expressions for 
 $\xi(r,t)$ which is the fluid displacement
vector and the other  metric potential $\delta \nu$. 

The generalized stochastic or probabilistic nature of these perturbations can
be concluded easily by denoting the statistical averages or expectation
with $< >_s $ for all the perturbed quantities and the source term. The
final statistical form of expressions read,
\begin{widetext}
\bea \label{eq:la1}
<\delta \lambda (r,t)>_s & =  & 4 \pi e^{\nu - \lambda} r \int_{r_1}^r 
\frac{e^{3 \lambda(r') - \nu(r')}}{[1 + \tilde{K}(\gamma_{r'})]} <\delta_s 
\epsilon(r')>_s  <e^{i {\omega}_{r'} t }>_s dr' 
\eea
The probabilistic nature of the amplitude and phase of oscillations of the 
source term are treated as independent random variables here. 
It can be easily seen that $ < \delta \lambda(r,t) >_s =0 $, since $ <\delta_s
\epsilon(r')>_s = 0 $ necessarily for the Langevin noise, as well as
 $ <e^{i \omega_{r'}t} >_s $ w.r.t the
random frequency distribution may vanish. Thus the  Langevin nature of the
source ensures the expectation of $<\delta \lambda(r,t) >_s$ and other 
perturbed variables to vanish.

Similarly,
 
\bea \label{eq:xif1}
<\xi (r,t)>_s & =  & 4 \pi (\nu'(r) + \lambda'(r)) e^{\nu - \lambda} r 
\int_{r_1}^r \frac{e^{3 \lambda(r') - \nu(r')}}{[1 + \tilde{K}(\gamma_{r'})]}
<\delta_s \epsilon(r')>_s< e^{i {\omega}_{r'} t }>_s dr' =0
\eea
and 
\bea \label{eq:nuf1}
<\delta \nu(r,t)>_s & = & \int Y_1(r',\gamma_{r'}) ( \int \frac{ e^{ 3 
\lambda(r'') - \nu(r'')}}{[1 + \tilde{K}(\gamma_{r''}) ]} <\delta_s 
\epsilon (r'')>_s <e^{i {\omega}_{r''} t }>_s dr'') dr' + \int Y_2(r', 
\gamma_{r'}) <\delta_s \epsilon ( r')>_s <e^{i {\omega}_{r'} t}>_s dr' \nonumber \\
& & + \int 4 \pi r' e^{2 \lambda(r')} <\delta_s p(r')>_s 
<e^{i {\omega}_{r'} t}>_s dr' = 0
\eea
where $<\xi(r,t)>_s $ and $ <\delta \nu(r,t)>_s $ also vanish as expected.
This is in accordance with  statistical fluctuations begin screened off
or vanishing in a large ensemble of $N$ particles, as they are supressed by 
a factor of $1/\sqrt{N} $ and thus physically negligible.  

However the two point correlations take the form,
\bea  \label{eq:twolamb}
<\delta_s \lambda(r,t) \delta_s \lambda^*(\bar{r},\bar{t})>_s & = & 8 \pi^2 
e^{\nu(r) + \nu(\bar{r}- \lambda(r) - \lambda(\bar{r}) } \int_{r_1}^r 
\int_{\bar{r_1}}^{\bar{r}} \frac{ e^{3(\lambda(r') 
 + \lambda(\bar{r'})) - \nu(r') - \nu(\bar{r'})}}{[1+ \tilde{K}(\gamma')][1+
 \tilde{K}(\bar{\gamma_r'})]} \nonumber \\
& &  < \delta_s \epsilon(r') \delta_s \epsilon(\bar{r'})>_s 
<e^{i(\omega_{r'} t - \omega_{\bar{r'}} \bar{t} ) }>_s dr' d\bar{r'}
\eea
and 
\bea \label{eq:twoxi}
< \xi(r,t) \xi^*(\bar{r},\bar{t})>_s & = & 8 \pi^2  (\nu'(r) + \lambda'(r))(\nu'(\bar{r}) + \lambda'(\bar{r}))
e^{\nu(r) + \nu(\bar{r}- \lambda(r) - \lambda(\bar{r}) } \int_{r_1}^r 
\int_{\bar{r_1}}^{\bar{r}} \frac{ e^{3(\lambda(r') 
 + \lambda(\bar{r'})) - \nu(r') - \nu(\bar{r'})}}{[1+ \tilde{K}(\gamma')][1+
 \tilde{K}(\bar{\gamma_r'})]} \nonumber \\
& &  < \delta_s \epsilon(r') \delta_s \epsilon(\bar{r'})>_s 
<e^{i(\omega_{r'} t - \omega_{\bar{r'}} \bar{t} ) }>_s dr' d\bar{r'}
\eea
\end{widetext}
a similar two point expression for $\delta \nu(r,t) $ can be obtained easily. 

In equation (\ref{eq:twolamb}) and (\ref{eq:twoxi}) the appearance of two 
point correlations, $< \delta_s \epsilon(r) \delta_s \epsilon(r')>_s,
<\delta_s p(r) \delta_s p(r') >_s$ and  $ <e^{i(\omega_{r'} t - 
\omega_{\bar{r'}} \bar{t} ) }>_s $ describe the two point generalized noise
 explicitly.
One can easily obtain expressions for
the point separated perturbations  with $t=\bar{t}, r \neq \bar{r} $ and 
$t \neq \bar{t}, r = \bar{r} $ as well as for the rms value with $ r = \bar{r}
, t = \bar{t} $ from equations (\ref{eq:twolamb} ) and (\ref{eq:twoxi}).  
 
In general from the above expressions one can see that the two point
 correlations need not necessarily be restricted to stationary conditions
with $t-\bar{t}$ dependency only. The individual perturbations 
equations (\ref{eq:la}) (\ref{eq:xif}) (\ref{eq:nuf}) are localized
and take into consideration the cumulative effect of the 
background sources along with the stationary condition for the induced effect
 via the $K(\gamma_r) $ factor. However the statistical average can only
be meaningful here for physical considerations given the generalized
stochastic nature. The two point correlations of these quantities do not
have any restriction as one can see from the expressions, of addressing only
stationary correlations and are valid for any separations in principle. 
However these valid separations will have limits due to the causality
conditions that appear in the results, nd will be useful for non-equilibrium
 studies. It is through this limit that
the intermediate sub-hydro mesoscopic scales can be defined and decided in the
dense matter.     
\section{Results and discussion}
We have shown that the statistical  perturbations $\xi, \delta \lambda, 
\delta\nu $ can be sourced by the random oscillations of pressure and energy 
density seeds present in the initial unperturbed relativistic 
star. As discussed earlier in detail, such a background noise 
 may be present due to various reasons even in the cold  dense matter
stars. In addition to the picture given for the consideration 
in this article, one can investigate more causes that may give rise to similar 
 trigger factors inside the matter of the star.  
The two point correlations of the induced 
perturbations (as above and similar ones for different 
configurations of matter content in various dense matter relativistic stars) 
act as the main framework for the statistical analysis of the 
 the dense fluid at a sub-hydro meso range. The appearance of $ \gamma_r $ as
 the random phase  factor 
which in general can have a real and a complex part decides the
evolution for the induced perturbations. It is worthwhile to note that
the phase factor of the sources  in terms of $\omega_r$ with its radial
dependence and that for the perturbations given in terms of $\gamma_r$ are
different. The details of the expressions give clear indication of the
cumulative effect that the source term has in inducing the
perturbed physical variables in the star.  
It is through the two point or higher correlations of the induced
 perturbations that  it will  enable us to obtain the non-equilibrium 
(dynamical) or near equilibrium properties and the new physical analysis of the
exotic dense matter. 
One can extend this theoretical framework later for non-linear
studies as well. However at present we restrict our developments to
linear perturbations only. 
\section{Conclusion and Further Directions} \label{sec:5}
With our formalism we have described the underlying phenomena of how induced
random perturbations can arise due to internal sources in a relativsitic star.
The growth and decay of such internally sourced perturbation
 is a further direction to explore. 
Formulating any disturbances in terms of the 
physical parameters  like pressure and energy density which are coarse grained
 effects  in the bulk of the matter is the first step in this
 approach.
Hence one is not flawed by such assumptions at this stage,  on the other hand,
 current research in dark matter also raises queries  over
 the partial quantum nature of exotic matter to be retained 
in  the macroscopic bulk fluid  description \cite{astron}. 

In this article we have given a  classical Einstein-
Langevin equation, based on a  linear response  relation
for matter in strong gravity regions. The  first simple model for radial
 perturbations of a spherically symmetric relativistic star has been
 solved in all details to confirm the mathematical and astrophysical aspects 
in this regard. Spherically symmetric star models and radial
perturbations are not just of importance as a  first exercise to 
do regarding any new developments, but they have been used consistently
for understanding stabilty of stars. The closed form analytical results
which give ideas about the basic physics or astrophysics, lay a
path for research directions towards realistic cases as well as observational
consequences. It is beginning with such idealized models for analytical
results which can be worked out neatly, that one
can progress further on building up deviations from symmetry and take as  
approximations for configurations of real and observational interests.  Thus
 from our results in this article one can see the connections and form of 
noise sources which are  very much of significance to collapse models.
This can affect the 
parameters that  decide the final outcome of a 
dynamical spherically symmetric astrophysical system. Even for equilibrium
configuration or near equilibrium configuration, the growth of such
 perturbations can be significant and a mode analysis of the spectrum
can lead to useful information of the deep interiors of the dense compact
 matter.  
Thus the significance of radial symmetry in star models is non-trivial, though
current focus remains on non-radial aspects for direct usefulness to
gravitational waves and their detection.  
We  will progress on the  theoretical developments towards
non-radial perturbations induced due to matter fluctuations or
disturbances. In the upcoming work we plan to explore several
configurations, including hot matter fluids with heat flux and anisotropies,
 stationary and non-stationary fluids etc. The main aim is to obtain  
equilibrium and non-equilibrium statistical properties  with information
pertaining to mesoscopic range inside the matter of relativistic stars. 
The known approaches towards studying the dense matter concentrate either on
the macroscopic  hydrodynamic scales about the nature of dense matter, or the 
astroparticle picture and the microscopics. The mesoscopic scale interest in
exploring the dense matter arises not just to fill in the scale gaps, but
also to  address the transition from the hydrodynamic scale down to the 
microscopic scales and see the connections and phases that the exotic 
matter can undergo. Thus our study would enhance  the physical picture of
interiors of the relativistic stars, the area which is gaining more importance
  for observational purposes as well.  
 
Our efforts at present are to establish the theory in full detail,  
 in this direction. One can expect that the observational consequences
may be varied for different configurations. Even for the electromagnetic 
spectra recieved from massive stars,it cannot be ruled out that  small 
parameter changes may be reflected in the data for such observations.     
However such interests can be addressed correctly and more precisely once the
basic rigorous theoretical framework is set up as initial stages in this
research program. 
Some models  may eventually require numerical methods to obtain solutions of 
the Einstein-Langevin equation. As of now we concentrate on
the simpler analytical closed form results that  can be obtained for a few 
cases to bring together the basic concepts from  general relativity,
astrophysics and statistical physics with new minor or major constructs 
and modifications, as may be required at each step. 
\section{Acknowledgements}
The author is thankful to Profs.,  Nils Andersson of helpful comments, Rajesh
 Nayak for initial discussions towards the overall theme
over a period of time, and to Profs, Deepak Dhar and
A.D Gangal for useful short discussions on basic concepts from statistical
physics. Part of the work done in this article was carried
out during the project funded by DST, India under grant no. 
DST-WoSA/PM100/2016 at IISER Pune, India. 

\end{document}